\documentclass[prx,aps,showpacs,twocolumn,
amsmath,amssymb,superscriptaddress,nofootinbib,longbibliography,nofootinbibjustification=justified,singlelinecheck=false]{revtex4-2}

\usepackage{graphicx}
\usepackage[table]{xcolor}
\usepackage[inkscapelatex=false]{svg}
\usepackage{braket}
\usepackage{multirow}
\usepackage{booktabs}
\usepackage{hyperref} 
\usepackage{cleveref}
\usepackage{listings}
\usepackage{color}
\usepackage[x11names,svgnames]{xcolor}
\hypersetup{
    hidelinks,
    colorlinks=true,
    breaklinks=true,
    citecolor=SteelBlue,
    filecolor=LimeGreen,
    linkcolor=MediumBlue,
    urlcolor=MediumPurple
}

\begin{document}

\title{Polynomial neural surrogates for designing photonic quantum experiments}

\author{Rohit Chaurasiya}
\email{rohit.chaurasiya@uni-jena.de}

\author{Xuemei Gu}
\email{xuemei.gu@uni-jena.de}
\affiliation{
Institute of Condensed Matter Theory and Optics, 
Friedrich Schiller University Jena, 
Max-Wien-Platz 1, 07743 Jena, Germany
}

\begin{abstract}

Physics simulators can support the discovery of quantum experiments by predicting the states generated by experimental configurations. When these simulators are computationally expensive, repeated simulator calls can limit the search for experiments that generate a desired quantum state. Here, we develop a physics-inspired polynomial neural surrogate for PyTheus, a graph-based quantum-optics simulator, to predict quantum states and use it to design quantum experiments. Its polynomial activations are motivated by the relation between graph perfect matchings and the resulting state amplitudes. We train separate surrogate models for four-, six-, and eight-photon systems and show that they achieve higher prediction accuracy with fewer trainable parameters than standard multilayer perceptrons. We then use the trained surrogates for inverse design of GHZ, W, and linear-cluster states. For the larger systems, the surrogates also enable faster inverse design than direct optimization with PyTheus. These results suggest that incorporating the underlying physics into neural surrogates can provide an efficient approach to quantum experiment design.

\end{abstract}

\maketitle

\section*{Introduction}
Physics simulators are widely used to predict the behaviour of experimental systems and increasingly serve as computational tools for experiment discovery~\cite{klimesch2026designing}. In automated design, candidate experimental configurations are evaluated and optimised according to a desired objective, often requiring many repeated calls to the underlying simulator~\cite{krenn2016automated,knott2016search,nichols2019designing,ruiz2023digital,rodriguez2024automated,krenn2025digital}. Such simulations provide a direct connection between the parameters describing an experiment and the physical quantities used to evaluate its performance. For computationally expensive simulators, however, repeated evaluations can become a bottleneck in the design process. Neural surrogate models provide an alternative by learning the input--output mapping of the simulator and replacing these evaluations with a faster learned approximation~\cite{peurifoy2018nanophotonic,ma2021deep, peano2021rapid,ruiz2026neural}. When the simulator has a known mathematical structure, this information can further provide useful prior knowledge for designing the surrogate architecture~\cite{greydanus2019hamiltonian,cranmer2020lagrangian,karniadakis2021physics,meng2025physics}.

PyTheus, which uses a graph-based quantum-optics simulator, provides such a structured setting for photonic quantum experiment design~\cite{ruiz2023digital}. It represents experimental configurations as weighted graphs and computes the corresponding quantum state from the perfect matchings of the graph~\cite{krenn2017quantum,gu2019quantum,gu2019quantum3, krenn2021conceptual}, where a perfect matching is a set of edges that contains each vertex exactly once. 
Previous work has used feed-forward neural networks to predict target-specific quantities directly from the graph weights, such as the fidelity with a fixed target state, and used the trained models to guide the design of quantum experiments toward that target~\cite{jaouni2024deep}. Rather than learning a particular property of the generated state, we learn the mapping from the graph weights to the quantum state itself. The surrogate predicts the state amplitudes, allowing different target states and design objectives to be introduced during the subsequent inverse-design stage without retraining the model. This graph-to-state mapping has a known polynomial structure, since each state amplitude is obtained by summing contributions from relevant perfect matchings, with each contribution given by a product of graph weights~\cite{krenn2017quantum, gu2019quantum, krenn2021conceptual}.

Motivated by this structure, we develop a physics-inspired polynomial neural surrogate for PyTheus that predicts the quantum state from the graph weights. We employ a polynomial neural network (PNN)~\cite{kileel2019expressive, kubjas2024geometry} with a single hidden layer, choosing the polynomial order to reflect the number of edges in each perfect matching. We train separate surrogate models for four-, six-, and eight-photon systems and show that they achieve lower prediction errors with substantially fewer trainable parameters than standard multilayer perceptrons. We further compare the Jacobian sensitivities~\cite{pizarroso2022neuralsens} of the learned and analytical mappings to examine how well the surrogates capture the dependence of the state amplitudes on the graph weights. Finally, we use the trained surrogates for inverse design of four-, six-, and eight-photon GHZ, W, and linear-cluster states. The resulting configurations achieve high fidelities when independently evaluated with PyTheus, and for the larger systems considered, the surrogates also enable faster inverse design than direct optimisation with PyTheus. These results demonstrate how the known algebraic structure of a physics simulator can guide the design of compact and reusable surrogates for photonic quantum experiment design.

\section*{From Weighted Graphs to Quantum States}
\label{sec:graph-representation}

PyTheus uses a weighted edge-coloured graph representation of quantum-optical experiments~\cite{krenn2017quantum, gu2019quantum, krenn2021conceptual,ruiz2023digital}. Each vertex represents an optical path to a detector, while an edge connecting two vertices $a$ and $b$ represents a correlation between the photons in paths $a$ and $b$. The edge weight indicates the amplitude of the photon pair, while the edge colours represent the internal modes of the photons in the corresponding paths, which can be encoded in polarisation, orbital angular momentum, time bins, frequency modes, or other photonic degrees of freedom~\cite{erhard2020advances}. Such a graph representation has a direct physical interpretation and can be translated into experimentally realizable photonic configurations, as demonstrated in several experiments~\cite{bao2023very,qian2023multiphoton,feng2023chip, wang2024entangling}.

\begin{figure}[!t]
    \centering
    \includegraphics[width=0.45\textwidth]{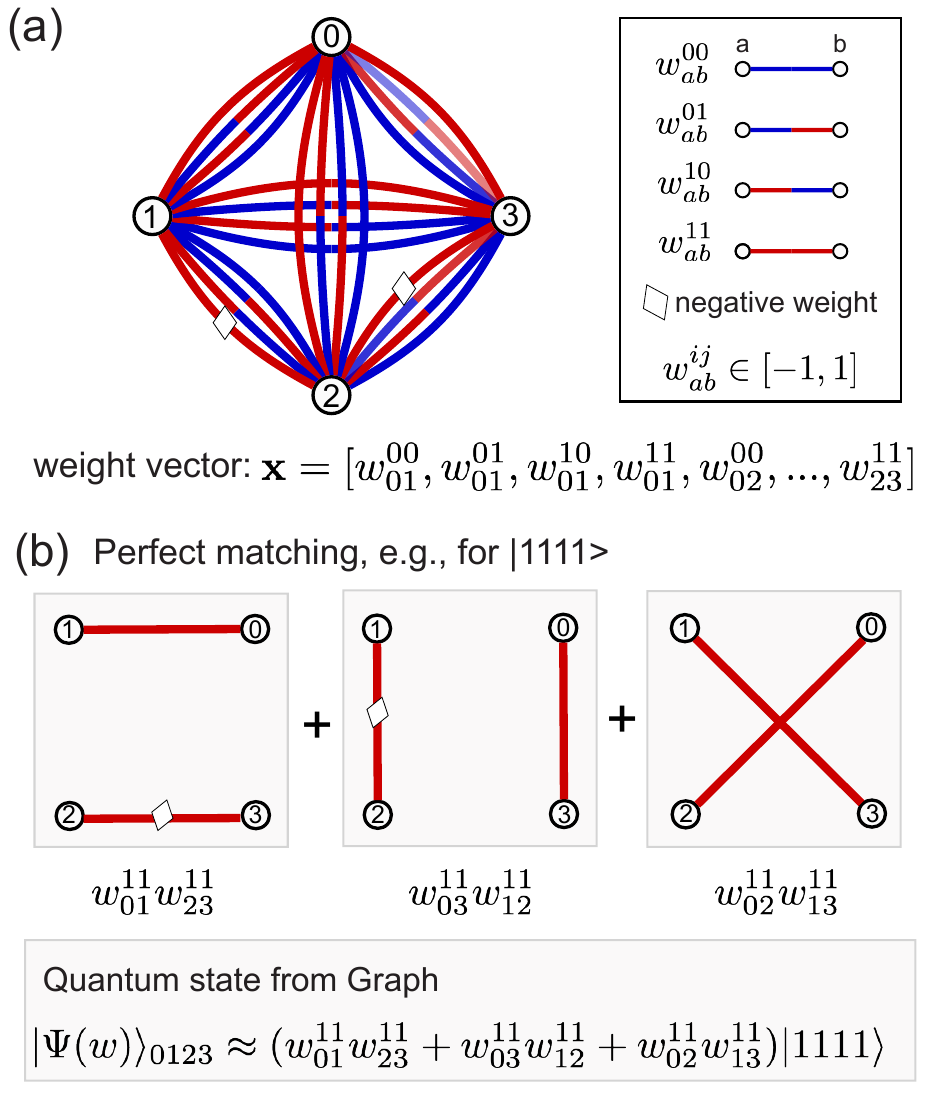}
   \caption{\textbf{Graph representation and perfect matching contributions.}
\textbf{(a)} Fully connected four-vertex graph. Blue and red colours represent the photon states $\ket{0}$ and $\ket{1}$, respectively. An edge joining vertices $a$ and $b$ with photon states $i$ and $j$ has weight $w_{ab}^{ij}\in[-1,1]$. The four colour combinations for each of the six vertex pairs give 24 weights, collected in the vector $\mathbf{x}$. Diamonds indicate negative weights.
\textbf{(b)} For $n=4$, three perfect matchings contribute to the $\ket{1111}$ amplitude. A perfect matching is a set of edges that covers every vertex exactly once. Summing the contributions from these three matchings gives the amplitude in the unnormalised output state:
\(\langle 1111|\Psi_{\mathrm{post}}\rangle
= w_{01}^{11}w_{23}^{11}
+ w_{03}^{11}w_{12}^{11}
+ w_{02}^{11}w_{13}^{11}\).}
    \label{fig:perfect-matching-decomposition}
\end{figure}

In Fig.~\ref{fig:perfect-matching-decomposition}(a), we consider a graph with four vertices, $V=\{0,1,2,3\}$, and two internal modes, labelled $0$ and $1$, represented by blue and red, respectively. For each pair of vertices, the two internal modes give rise to four possible colour combinations. We denote the edge weight by $w_{ab}^{ij}$, where $a$ and $b$ label the vertices (or optical paths), and $i$ and $j$ denote the internal modes of the two photons. The fully connected graph in Fig.~\ref{fig:perfect-matching-decomposition}(a) therefore contains 24 edges with real-valued weights in $[-1,1]$. More generally, a fully connected graph with $n$ vertices and two internal modes contains $2n(n-1)$ weighted edges. The quantum state generated by the graph can be expressed in terms of the edge weights~\cite{krenn2021conceptual, ruiz2023digital},
\begin{align}
\Phi(\mathbf{w})
=
\sum_{m=0}^{\infty}
\frac{1}{m!}
\left(
\sum_{e\in E(G)}
w(e)\,
\hat{x}^{\dagger}(e)
\hat{y}^{\dagger}(e)
+\mathrm{h.c.}
\right)^{m}
\label{eq:graph_weight_function}
\end{align}
where $\hat{x}^{\dagger}(e)$ and $\hat{y}^{\dagger}(e)$ create photons at the two vertices $x$ and $y$ connected by edge $e$, and $\mathrm{h.c.}$ denotes the Hermitian conjugate. Applying $\Phi(\mathbf{w})$ to the vacuum yields the generated photonic state.

Experimentally, we often post-select the quantum state using $n$-fold coincidence detection, corresponding to the simultaneous detection of one photon in each of the $n$ output paths. Since each edge corresponds to a photon pair, an $n$-photon event contains $n/2$ edges and therefore requires even $n$. Accordingly, the $n$-photon contribution arises from the $m=n/2$ term in Eq.~\eqref{eq:graph_weight_function},
\begin{align}
\left(
\sum_{e\in E(G)}
w(e)\,
\hat{x}^{\dagger}(e)
\hat{y}^{\dagger}(e)
\right)^{n/2}.
\label{eq:n_photon_expansion}
\end{align}
Expanding this term gives products of $n/2$ edge contributions. The coincidence condition selects only those products for which each output path occurs exactly once. Thus, the corresponding edges share no vertices and cover every vertex exactly once. These edge sets are the perfect matchings of the graph.

Let $M$ denote a perfect matching. We denote an output state by
$\ket{\mathbf{s}}=\ket{s_0s_1\ldots s_{n-1}}$, where $s_i$ specifies the mode of the photon in output path $i$. For example, $\ket{0101}$ represents the mode encoding $0,1,0,1$ across the four output paths. The weight of a perfect matching is the product of its edge weights,
\begin{equation}
W(M)=\prod_{e\in M} w(e).
\end{equation}
Different perfect matchings can generate the same output state $\ket{\mathbf{s}}$. Its unnormalized amplitude is therefore obtained by summing the weights of all perfect matchings with the corresponding edge colours. The post-selected state can thus be written as
\begin{equation}
\ket{\Psi_{\mathrm{post}}}
=
\sum_{\mathbf{s}}
\left(
\sum_{M\in\operatorname{PM}(G_{\mathbf{s}})}
W(M)
\right)
\ket{\mathbf{s}},
\label{eq:perfect_matching_amplitude}
\end{equation}
where $\operatorname{PM}(G_{\mathbf{s}})$ denotes the set of perfect matchings associated with $\ket{\mathbf{s}}$. For further details, we refer the reader to the PyTheus work~\cite{ruiz2023digital}.

This construction also reveals the polynomial structure of the graph-to-state mapping. Each term in the $m=n/2$ expansion contains a product of $n/2$ graph weights, while coincidence post-selection selects the subset corresponding to perfect matchings. The amplitude of each output state is therefore a sum of selected products of $n/2$ graph weights. This structure motivates the polynomial activation used in the neural surrogate introduced in the following section.

\section*{Surrogate Training}
\label{sec:surrogate}
Having established the polynomial dependence of the output amplitudes on the graph weights, we investigate whether this structure can be used to construct an efficient surrogate for predicting the complete output state. We compare the PNN with a standard multilayer perceptron (MLP) and evaluate their prediction accuracy, model size, and computational cost. We further examine whether the PNN captures how the output amplitudes change as the graph weights vary. 

\begin{figure}[!t]
    \centering
\includegraphics[width=0.49\textwidth]{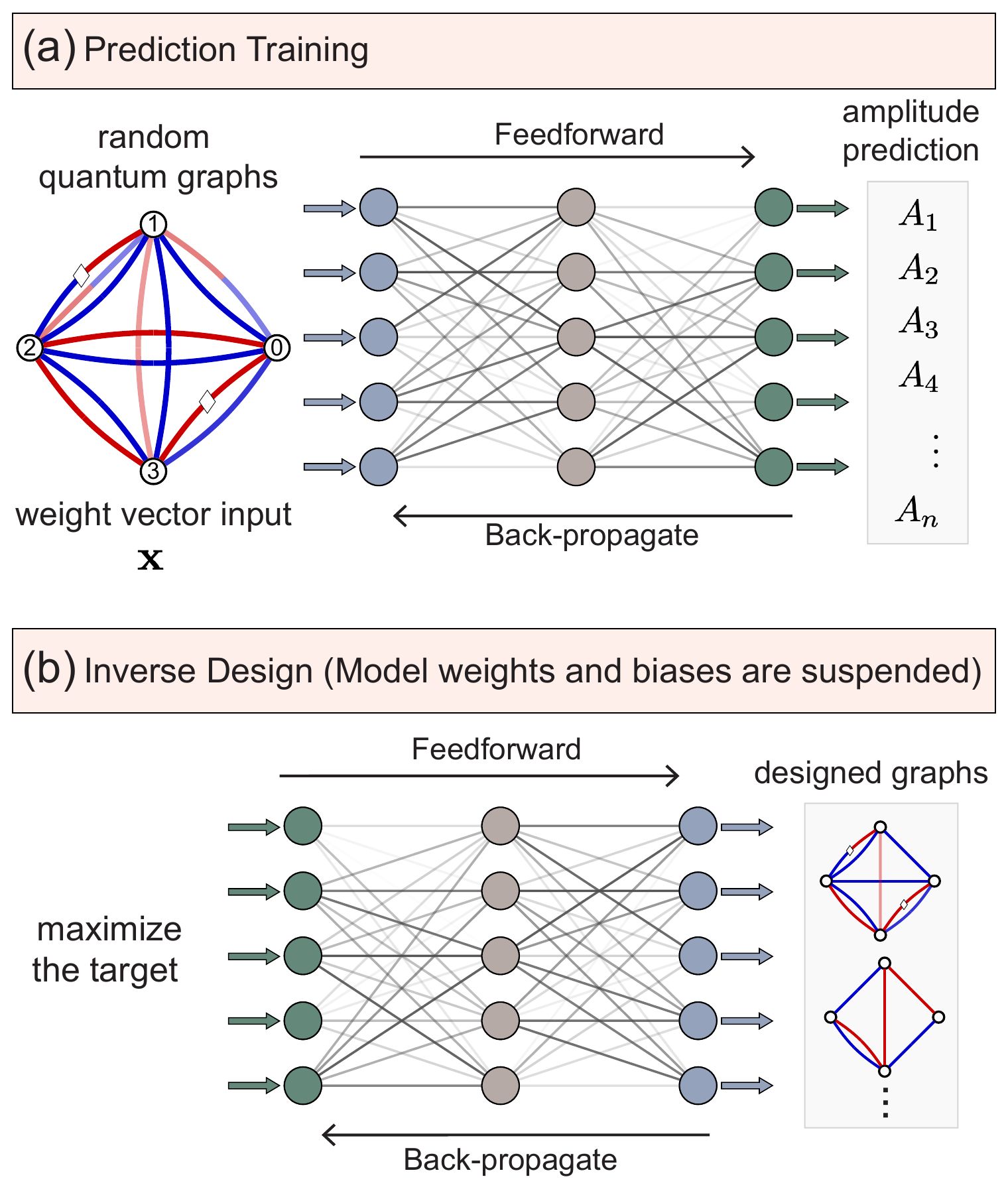}
    \caption{\textbf{Surrogate training and inverse design.} \textbf{(a)} Surrogate training. A weighted graph, encoded by the graph-weight vector $\mathbf{x}$, is passed through the neural network to predict the corresponding quantum-state amplitudes $\ket{\Psi_{\mathrm{NN}}}$. 
    \textbf{(b)} Surrogate-assisted inverse design. A randomly initialised graph-weight vector $\mathbf{x}$ is passed through the trained model, with its weights and biases frozen. The graph weights are iteratively updated to maximize the target evaluated from the predicted state until the stopping criterion is reached.}
    \label{fig:inverse-design-workflow}
\end{figure}

The surrogate-training procedure is shown in Fig.~\ref{fig:inverse-design-workflow}(a). We first consider the $n=4$ case, for which the graph contains $24$ weighted edges. Their weights form the input vector $\mathbf{x}=(x_1,\ldots,x_{24})$, with each $x_i$ independently sampled from a uniform distribution on $[-1,1]$. For each sampled graph, PyTheus computes the corresponding $16$ real, unnormalised output-state amplitudes, collected in $\boldsymbol{A}$ and used as the supervised target. We generate $20$ million such input--output pairs and divide them into training, validation, and test sets containing $80\%$, $10\%$, and $10\%$ of the data, corresponding to $16$, $2$, and $2$ million samples, respectively.

We then train two surrogate models on the same dataset, a standard MLP and the physics-guided PNN. The MLP serves as a baseline and consists of three fully connected hidden layers with GELU activations~\cite{hendrycks2016gaussian}, followed by a linear output layer. The PNN incorporates the polynomial structure of the graph-to-state mapping established in Section~\ref{sec:graph-representation}. For $n=4$, each perfect-matching contribution is quadratic in the graph weights, motivating a single fully connected hidden layer with a quadratic power activation. More generally, for even $n$, the PNN predicts the output amplitudes as
\begin{equation}
\hat{\boldsymbol{A}}
=
\mathbf{V}\,
\sigma_{n/2}(\mathbf{U}\mathbf{x}+\mathbf{b})
+\mathbf{c},
\label{eq:PNN-architecture}
\end{equation}
where $\mathbf{U}$ and $\mathbf{V}$ are trainable weight matrices, $\mathbf{b}$ and $\mathbf{c}$ are biases, and $\sigma_{n/2}(z)=z^{n/2}$ is applied element-wise. 

\begin{table*}[!t]
\caption{
Comparison of the neural-network architectures used for surrogate training.
The best epoch corresponds to the minimum validation loss, while epochs trained
denotes the total number of epochs before termination.
}
\label{tab:architecture-comparison}
\centering
\renewcommand{\arraystretch}{1.08}
\begin{tabular*}{\textwidth}{
@{\extracolsep{\fill}}
c l c c c c c c c r
}
\toprule
$n$
& Model
& Layers
& Neurons
& Activation
& Best epoch
& Epochs trained
& Best val. loss
& Training time
& Parameters \\
\midrule

\multirow{2}{*}{$4$}
& PNN & 1 & 400 & $z^2$ & 2506 & 4506
& $9.13\times10^{-7}$ & $0.93\,\mathrm{h}$ & 16,416 \\
& FNN & 3 & 400 & GELU & 19819 & 20000
& $2.05\times10^{-5}$ & $23.12\,\mathrm{h}$ & 337,216 \\
\midrule

\multirow{2}{*}{$6$}
& PNN & 1 & 2000 & $z^3$ & 2497 & 4497
& $3.90\times10^{-6}$ & $3.81\,\mathrm{h}$ & 250,064 \\
& FNN & 3 & 2000 & GELU & 19975 & 20000
& $2.21\times10^{-4}$ & $292.63\,\mathrm{h}$ & 8,254,064 \\
\midrule

\multirow{2}{*}{$8$}
& PNN & 1 & 15000 & $z^4$ & 2984 & 4984
& $5.93\times10^{-5}$ & $55.6\,\mathrm{h}$ & 5,535,256 \\
& FNN & 3 & 8000 & GELU & 3119 & 5119
& $2.85\times10^{-1}$ & $1021.08\,\mathrm{h}$ & 130,968,256 \\
\bottomrule
\end{tabular*}
\end{table*}
\begin{figure*}[t]
    \centering
\includegraphics[width=1\textwidth]{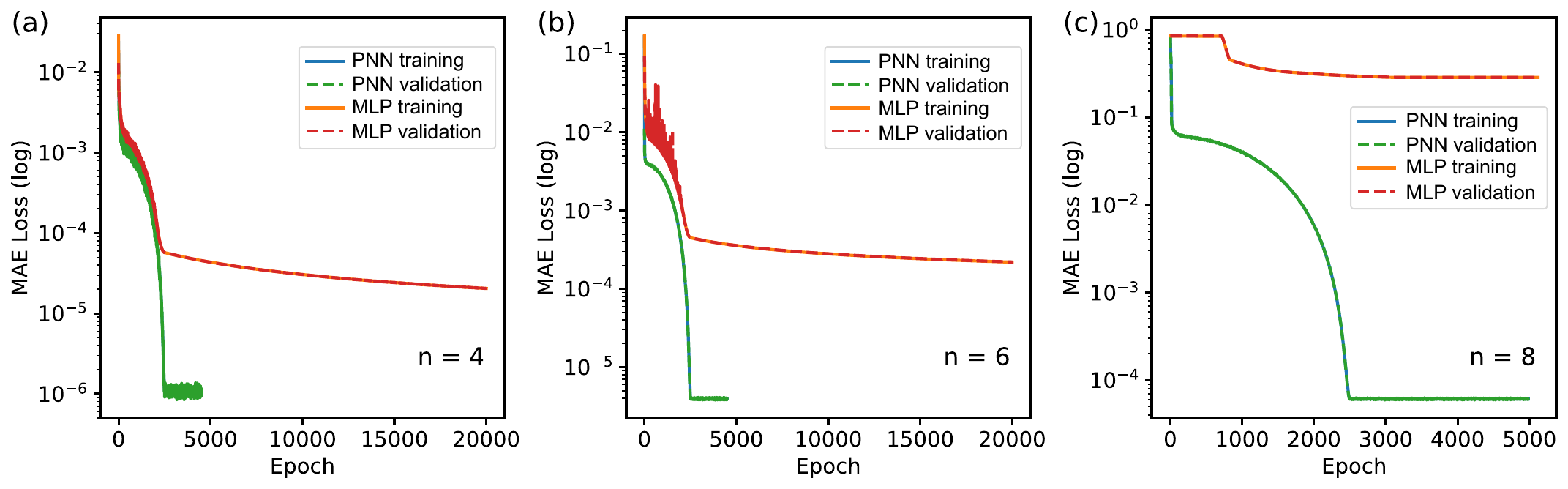}
\caption{
Training and validation loss curves for the PNN and FNN surrogate models for $n=4$, $6$, and $8$. The mean absolute error (MAE) between the predicted and PyTheus-generated unnormalised output states is shown on a logarithmic scale. Across all system sizes, the PNN achieves a lower validation loss than the MLP. For $n=8$, MLP training stopped after $2000$ epochs without further improvement. The best validation losses for all architectures and system sizes are reported in Table~\ref{tab:architecture-comparison}.
} \label{fig:hnn-fnn-training-comparison}
\end{figure*}

The power activation matches the polynomial degree of the graph-to-state mapping, while the PNN does not explicitly encode the individual perfect matchings. Instead, the combinations of graph weights required to reproduce the output amplitudes are learned from the training data.

Both models are trained by minimizing the mean absolute error (MAE) between the predicted and PyTheus-generated output amplitudes. We use AdamW with a batch size of $5000$ and an initial learning rate of $10^{-3}$, followed by cosine decay. The model with the lowest validation MAE is retained for test evaluation and subsequent inverse design. Training is terminated if the validation MAE does not improve for $2000$ epochs or when $20\,000$ epochs are reached.

The same procedure is applied to $n=6$ and $n=8$. Following the polynomial degree of the corresponding perfect-matching contributions, the PNN uses cubic and quartic power activations, respectively. For each case, we also generate $20$ million input--output pairs with the same training, validation, and test split. The model widths, activation functions, parameter counts, and training statistics are summarized in Table~\ref{tab:architecture-comparison}.

Fig.~\ref{fig:hnn-fnn-training-comparison} shows the training and validation MAE for the PNN and MLP across all three system sizes. Under the same training protocol, the PNN consistently achieves lower validation error with substantially fewer trainable parameters and shorter training times, as summarized in Table~\ref{tab:architecture-comparison}. The difference is particularly pronounced for $n=8$, where the PNN reaches a validation MAE of $5.93\times10^{-5}$, compared with $2.85\times10^{-1}$ for the MLP. For $n=8$, we additionally tested larger MLP architectures and a larger training dataset, but observed no further improvement. 

\begin{figure*}[!t]
    \centering
    \includegraphics[width=1\textwidth]{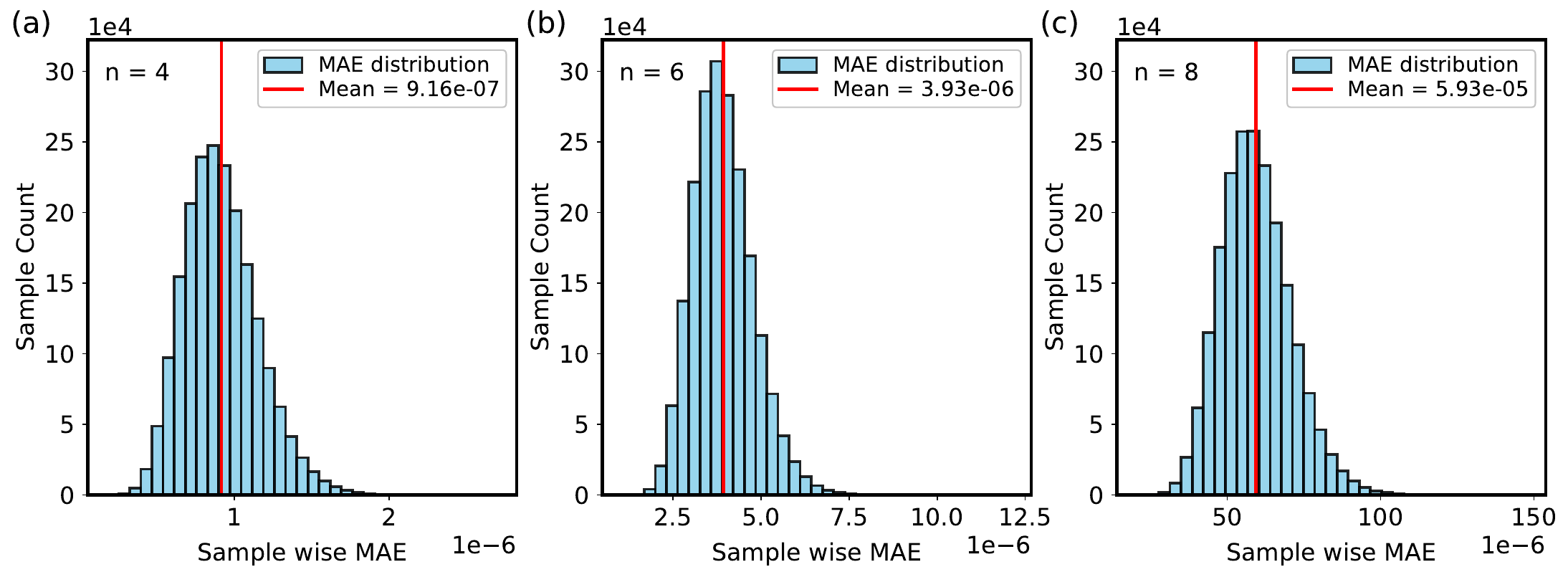}
    \caption{\textbf{Test-set prediction errors of the trained PNN surrogates.} Distributions of the sample-wise mean absolute error (MAE) for the \textbf{(a)} $n=4$, \textbf{(b)} $n=6$, and \textbf{(c)} $n=8$ systems. Each histogram contains all two million samples from the corresponding held-out test set. For each sample, the MAE is the mean absolute difference between the PNN-predicted and PyTheus-generated unnormalised output-state amplitudes over all output components. The red vertical line indicates the mean sample-wise MAE for each system size. The horizontal axes are shown in units of $10^{-6}$.}
    \label{fig:test-mae-distributions}
\end{figure*}

The trained PNNs also show low prediction errors on the test sets. The mean test MAEs are $9.162\times10^{-7}$, $3.93\times10^{-6}$, and $5.934\times10^{-5}$ for $n=4$, $6$, and $8$, respectively. These values are close to the validation errors, indicating consistent prediction accuracy on unseen samples drawn from the same distribution. The sample-wise test-error distributions are shown in Fig.~\ref{fig:test-mae-distributions}.

The results above show that the PNN accurately predicts the output amplitudes. For gradient-based inverse design, however, it should also capture how these amplitudes change with the graph weights, since these derivatives determine the optimization direction~\cite{olearyroseberry2024derivative,luo2025efficient}. We therefore compare its Jacobian with the analytical perfect-matching result,
\begin{equation}
J_{ij}^{(m)}(\mathbf{x})
=
\frac{\partial A_i^{(m)}(\mathbf{x})}{\partial x_j},
\qquad
m\in\{\mathrm{PNN},\mathrm{PM}\},
\label{eq:jacobian}
\end{equation}
where $J_{ij}^{(m)}$ describes how output amplitude $i$ changes with graph weight $j$. 

We evaluate these Jacobians on the same set of $N_J=1000$ sampled graphs and average the absolute value of each Jacobian element,
\begin{equation}
S_{ij}^{(m)}
=
\frac{1}{N_J}
\sum_{k=1}^{N_J}
\left|
J_{ij}^{(m)}(\mathbf{x}^{(k)})
\right|,
\label{eq:mean-jacobian}
\end{equation}
where the absolute value prevents derivatives with opposite signs from cancelling during averaging. Each row of $S^{(m)}$ represents one output amplitude, while each column corresponds to one graph weight.

\begin{figure*}[!t]
    \centering
\includegraphics[width=\textwidth]{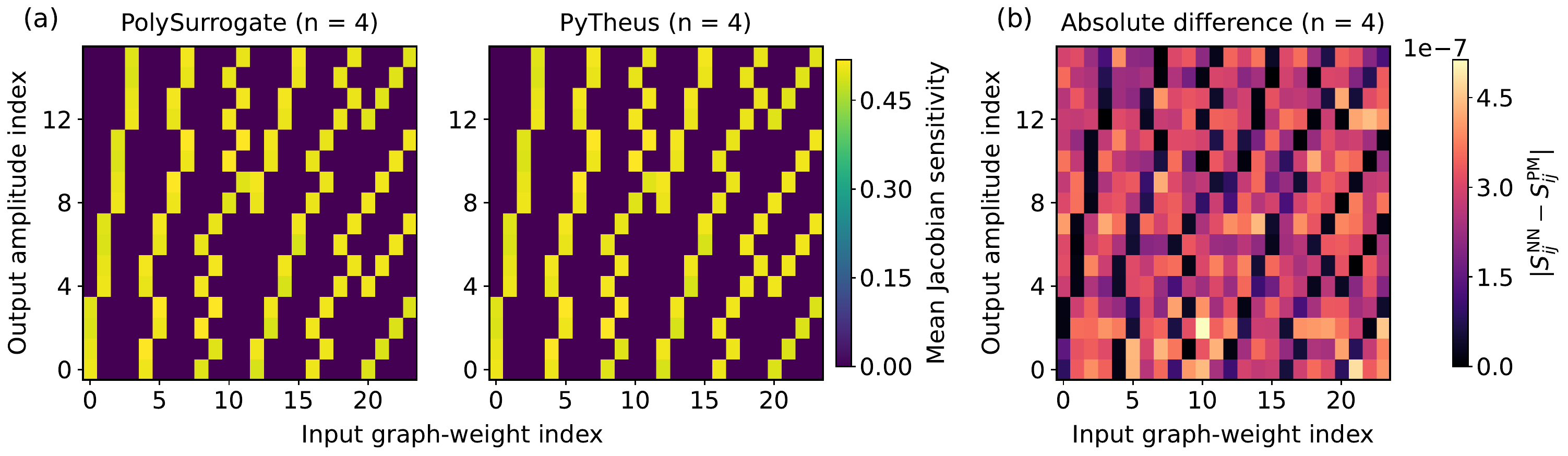}
    \caption{\textbf{Jacobian sensitivity comparison for $n=4$.} \textbf{(a)} Mean absolute sensitivities from the PNN (left) and analytical perfect-matching calculation (right), averaged over $1000$ sampled graphs. \textbf{(b)} Absolute element-wise difference between the two sensitivity matrices. The similar patterns and small differences show that the PNN reproduces the sensitivity of the output amplitudes to variations in the graph weights}
\label{fig:jacobian-sensitivity-4}
\end{figure*}

The PNN and analytical sensitivity matrices agree closely for all three system sizes, as illustrated for $n=4$ in Fig.~\ref{fig:jacobian-sensitivity-4}. The Pearson correlation between $S^{(\mathrm{NN})}$ and $S^{(\mathrm{PM})}$ exceeds $0.99999$ for each $n$, indicating that the two methods reproduce nearly the same pattern of dependencies between graph weights and output amplitudes. The mean element-wise absolute differences are $2.398\times10^{-7}$, $8.475\times10^{-7}$, and $9.918\times10^{-6}$ for $n=4$, $6$, and $8$, respectively, while the largest element-wise differences are $5.132\times10^{-7}$, $3.893\times10^{-6}$, and $1.88\times10^{-5}$.

For $n=4$, as an example, the analytical amplitude of the quantum state $\ket{0000}$ is
\begin{equation}
A_{0000}
=
w_{01}^{00}w_{23}^{00}
+
w_{02}^{00}w_{13}^{00}
+
w_{03}^{00}w_{12}^{00},
\end{equation}
and therefore depends only on the six edge weights
$\{w_{01}^{00},w_{02}^{00},w_{03}^{00},w_{12}^{00},w_{13}^{00},w_{23}^{00}\}$,
corresponding to weights $\{w_0,w_4,w_8,w_{12},w_{16},w_{20}\}$ in the graph-weight vector. Its derivatives with respect to these weights are determined by their partners in the corresponding perfect matchings; for example,
\begin{equation}
\frac{\partial A_{0000}}{\partial w_{01}^{00}}
=
w_{23}^{00},
\quad
\frac{\partial A_{0000}}{\partial w_{02}^{00}}
=
w_{13}^{00},
\quad
\frac{\partial A_{0000}}{\partial w_{03}^{00}}
=
w_{12}^{00},
\end{equation}
while derivatives with respect to all other graph weights vanish. The analytical sensitivity matrix therefore has a sparse and structured pattern determined by the perfect matchings contributing to each output amplitude. As shown in Fig.~\ref{fig:jacobian-sensitivity-4}, the PNN reproduces the same pattern, with negligible sensitivity to graph weights that do not contribute to a given amplitude and closely matching sensitivities for those that do. 

The sensitivity maps for $n=6$ and $n=8$ are provided in the Appendix \ref{app:perfect-matching-derivation} and show similarly close agreement. These results indicate that the PNN learns not only the output amplitudes but also the dependence of those amplitudes on the relevant graph weights, supporting its use for gradient-based inverse design.
\begin{figure*}[!t]
\centering
\includegraphics[width=1\textwidth]{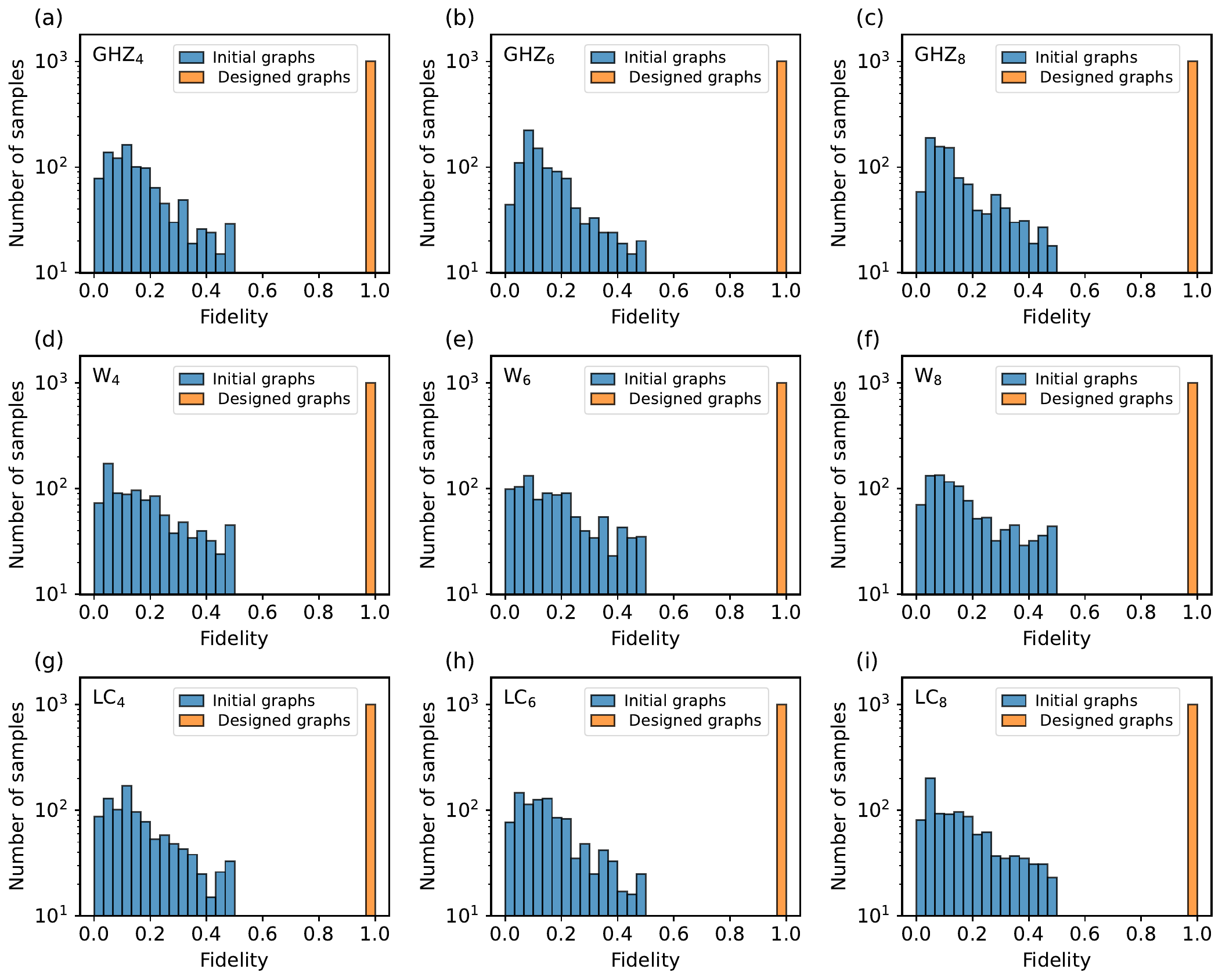}
\caption{\textbf{Surrogate-assisted inverse design.}
Distributions of the initial and final fidelities for \(1000\) random graphs for each target state and system size, with each graph optimized independently using the PNN surrogate. Blue and orange indicate the initial and final fidelities, respectively, with counts shown on a logarithmic scale.
}\label{fig:initial-final-fidelity-histograms}
\end{figure*}

\section*{From State Prediction to Inverse Design}
\label{sec:inverse-design}
We employ the trained PNN for inverse design by fixing the network parameters and optimizing the input graph weights to generate a desired target state, as illustrated in Fig.~\ref{fig:inverse-design-workflow}(b). For each system size $n=4$, $6$, and $8$, the corresponding PNN is used to design GHZ, W, and linear-cluster (LC) states. We perform $1000$ independent searches for each target and system size, giving $9000$ searches in total.

Different quantum experiment-design tasks may target different properties of the output state. For the state-generation tasks considered here, we use the state fidelity as the optimization objective as it directly quantifies the overlap between the predicted and target states. For the normalized PNN-predicted state $\ket{\hat{\psi}(\mathbf{x})}$, the fidelity is
\begin{equation}
\mathcal{F}(\mathbf{x})
=
\left|
\braket{\psi_{\mathrm{tar}}|\hat{\psi}(\mathbf{x})}
\right|^2.
\label{eq:inverse-design-fidelity}
\end{equation}
To obtain high-fidelity solutions while encouraging sparse graph configurations, we minimize
\begin{equation}
\mathcal{L}_{\mathrm{optim}}(\mathbf{x})
=
1-\mathcal{F}(\mathbf{x})
+\lambda_1\|\mathbf{x}\|_1,
\qquad
\lambda_1=10^{-3},
\label{optimiser_loss}
\end{equation}
where the $L_1$ term encourages small graph weights to approach zero.

For each target state, we first generate random initial graphs by independently sampling all graph weights from $[-1,1]$. We evaluate the PNN-predicted fidelity of each graph with respect to the target GHZ, W, or LC state and retain only graphs with an initial fidelity below $0.5$. These low-fidelity graphs are then used as the starting points for inverse design, where the graph weights are optimized to increase the fidelity towards unity.

The graph weights are optimized using Adam with an initial learning rate of $10^{-2}$, cosine-decayed to $10^{-6}$ over $100\,000$ steps. After each update, the weights are clipped to $[-1,1]$. Optimization stops when the predicted fidelity reaches $0.99999$ or when the maximum number of $100\,000$ steps is reached. During optimization, the increase in fidelity is evaluated every $1000$ steps. If the increase is smaller than $10^{-4}$, Gaussian noise $\boldsymbol{\epsilon}\sim\mathcal{N}(\mathbf{0},\mathbf{I})$ is added to the graph weights with
$
\mathbf{x}
\leftarrow
\operatorname{clip}
\left(
\mathbf{x}+\sigma\boldsymbol{\epsilon},
-1,1
\right)
$
and the Adam state is reset. The noise scale $\sigma$ increases after each perturbation as $0.01$, $0.05$, $0.10$, $0.20$, $0.30$, $0.50$, $0.70$, and $1.00$, remaining at $1.00$ thereafter.

\begin{table}[!t]
\centering
\caption{
PyTheus verification of the graphs designed using the PNN surrogate.
Each row contains $1000$ graphs. The last two columns report the fraction
with PyTheus-verified fidelities above the indicated thresholds. Samples with fidelities slightly below $0.99999$ correspond to runs that reach the maximum optimization step limit.}
\label{tab:pytheus-verification}
\setlength{\tabcolsep}{4pt}
\renewcommand{\arraystretch}{1.2}
\begin{tabular}{l c c c c c}
\toprule
Target
& $n$
& $\overline{F}_{\mathrm{PNN}}$
& $\overline{F}_{\mathrm{PyTheus}}$
& \multicolumn{2}{c}{PyTheus-verified fraction} \\
\cline{5-6}
&
&
&
&
$F\geq 0.9999$
&
$F\geq 0.99999$ \\
\hline

\multirow{3}{*}{GHZ}
& 4 & 0.999991 & 0.999991 & 100\% & 84.4\% \\
& 6 & 0.999991 & 0.999991 & 100\% & 85.3\% \\
& 8 & 0.999991 & 0.999991 & 100\% & 95.1\% \\
 
\multirow{3}{*}{W}
& 4 & 0.999990 & 0.999990 & 100\% & 58.6\% \\
& 6 & 0.999991 & 0.999991 & 100\% & 83.2\% \\
& 8 & 0.999991 & 0.999991 & 100\% & 85.0\% \\
 
\multirow{3}{*}{LC}
& 4 & 0.999990 & 0.999990 & 100\% & 82.2\% \\
& 6 & 0.999991 & 0.999991 & 100\% & 84.8\% \\
& 8 & 0.999991 & 0.999992 & 100\% & 93.0\% \\
\bottomrule
\end{tabular}
\end{table}
The resulting fidelity distributions are shown in Fig.~\ref{fig:initial-final-fidelity-histograms}. Across all targets and system sizes, PNN-assisted inverse design increases the fidelities from the low-fidelity initial graphs to values concentrated near unity. We then independently verify all $9000$ optimized graphs using PyTheus, and the verified fidelities closely agree with the corresponding PNN-predicted values, as summarized in Table~\ref{tab:pytheus-verification}. All verified graphs satisfy $F\geq0.9999$, and many also reach the stricter threshold $F\geq0.99999$.

We further compare the inverse-design runtimes in Table~\ref{tab:runtime-comparison}. For $n=4$, direct PyTheus optimization is generally faster on CPU, whereas for the larger $n=6$ and $n=8$ systems, surrogate-assisted optimization becomes substantially faster. GPU execution further reduces the PNN runtime, particularly for $n=8$. These measurements include only the post-training optimization stage and exclude the costs of dataset generation and surrogate training. The observed speedup should therefore be interpreted as a reduction in the cost of repeated inverse-design calculations after training, rather than as a reduction in the total end-to-end computational cost.
\begin{table}[!t]
\centering
\caption{
Inverse-design runtimes over $50$ runs, reported as mean $\pm$ standard error of the mean. PyTheus was evaluated on CPU, as GPU acceleration is not available in PyTheus, while the PNN was evaluated on both CPU and GPU.
}
\label{tab:runtime-comparison}
\setlength{\tabcolsep}{4pt}
\renewcommand{\arraystretch}{1.2}
\begin{tabular}{l c c c c}
\toprule
Target 
& $n$
& \shortstack{PyTheus\\CPU (s)}
& \shortstack{PNN\\CPU (s)}
& \shortstack{PNN\\GPU (s)}
\\
\hline

\multirow{3}{*}{GHZ}
& 4 & $1.44 \pm 0.81$       & $2.88 \pm 0.27$     & $1.28 \pm 0.14$ \\
& 6 & $56.56 \pm 71.79$     & $3.65 \pm 1.68$     & $2.87 \pm 2.88$ \\
& 8 & $2754.79 \pm 1552.90$ & $12.56 \pm 11.13$   & $5.86 \pm 5.04$ \\

\multirow{3}{*}{W}
& 4 & $0.71 \pm 0.38$       & $2.89 \pm 0.30$     & $1.71 \pm 0.30$ \\
& 6 & $20.41 \pm 5.06$      & $4.96 \pm 3.13$     & $5.81 \pm 6.38$ \\
& 8 & $1523.95 \pm 166.61$  & $34.96 \pm 27.05$   & $13.13 \pm 11.84$ \\

\multirow{3}{*}{LC}
& 4 & $0.90 \pm 0.24$       & $2.76 \pm 0.48$     & $1.47 \pm 0.40$ \\
& 6 & $117.02 \pm 79.47$    & $4.94 \pm 3.35$     & $6.07 \pm 6.98$ \\
& 8 & $10349.96 \pm 7092.97$& $46.61 \pm 38.29$   & $15.25 \pm 14.35$ \\
\bottomrule
\end{tabular}
\end{table}

\section*{Discussion and Outlook}
Our results show that the graph-to-state mapping can be accurately approximated by a PNN whose activation degree matches the known polynomial degree of the mapping. For the systems studied, this physics-guided architecture achieves lower prediction errors with substantially fewer parameters than a standard MLP. Predicting the complete output state also allows one model per system size to be reused for different inverse-design objectives without retraining. These results suggest that known algebraic structure can provide a useful inductive bias for quantum-experiment design~\cite{kileel2019expressive,kubjas2024geometry}.

The physics incorporated into the PNN is nevertheless limited. The activation degree $p=n/2$ reflects the polynomial degree of the graph-to-state mapping, but the network does not explicitly encode the perfect matchings or the exact monomial structure of the amplitudes. It can therefore represent polynomial terms that are absent from the exact mapping. Future models could incorporate more of the polynomial structure using architectures that explicitly parameterize high-order polynomial interactions~\cite{chrysos2021deep}, or by restricting the allowed monomials according to the underlying perfect-matching structure.

Our work also has several limitations in scope and scalability. We consider only two internal modes per photon and four-, six-, and eight-photon systems, with a separate surrogate for each system size. The larger hidden width required for the eight-photon case suggests that matching the polynomial degree alone may not scale to substantially larger systems. In addition, predicting the complete state retains the exponential growth of the output dimension. One promising direction is therefore to develop a unified model across system sizes, potentially using set-based or attention-based architectures capable of processing variable-size inputs~\cite{vaswani2017attention,lee2019set}. Such a model could learn structures shared across photon numbers rather than requiring an independent surrogate for each size. Extending to larger systems may also require compressed or task-relevant state representations, or models that predict selected amplitudes or observables instead of the complete state. Another interesting direction is to extend the framework from real- to complex-valued graph weights, potentially using complex-valued neural architectures~\cite{lee2022complex}. These directions could broaden the applicability of neural surrogates for photonic quantum experiment design.

\section*{Computational Resources}
All numerical experiments were performed on the Draco high-performance computing cluster at Friedrich Schiller University Jena. Model training and surrogate-assisted inverse design were carried out on a NVIDIA A100 GPU, while CPU runtime benchmarks used Intel Xeon Platinum 8360Y processors. The models were implemented in JAX/Flax with automatic differentiation.

\section*{Data and Code Availability}
The code for data generation, surrogate-model training, and inverse-design experiments is available at \url{https://github.com/AI-Quantum-Lab/PolySurrogate}.

\begin{acknowledgements}
X.G. acknowledges support from the Deutsche Forschungsgemeinschaft (DFG, German Research Foundation) -- Project-ID 398816777 -- SFB 1375.
\end{acknowledgements}


\bibliographystyle{apsrev4-2}
\bibliography{ref}

@article{klimesch2026designing,
  title={Designing physics experiments with artificial intelligence},
  author={Klimesch, Jonathan and Arlt, S{\"o}ren and Ruiz-Gonzalez, Carlos and Rodr{\'\i}guez, Carla and Gu, Xuemei and Haslinger, Philipp and Vischia, Pietro and Haack, Christian and Drori, Yehonathan and Adhikari, Rana and others},
  journal={Nature},
  volume={657},
  number={8130},
  pages={47--58},
  year={2026},
  publisher={Nature Publishing Group UK London},
  url={https://doi.org/10.1038/s41586-026-10898-6}
}

@article{krenn2025digital,
  title={Digital discovery of interferometric gravitational wave detectors},
  author={Krenn, Mario and Drori, Yehonathan and Adhikari, Rana X},
  journal={Physical Review X},
  volume={15},
  number={2},
  pages={021012},
  year={2025},
  publisher={APS},
  url={https://doi.org/10.1103/PhysRevX.15.021012}
}

@article{rodriguez2024automated,
  title={Automated discovery of experimental designs in super-resolution microscopy with XLuminA},
  author={Rodr{\'\i}guez, Carla and Arlt, S{\"o}ren and M{\"o}ckl, Leonhard and Krenn, Mario},
  journal={Nature Communications},
  volume={15},
  number={1},
  pages={10658},
  year={2024},
  publisher={Nature Publishing Group UK London},
  url={https://doi.org/10.1038/s41467-024-54696-y}
}

@article{krenn2016automated,
  title={Automated search for new quantum experiments},
  author={Krenn, Mario and Malik, Mehul and Fickler, Robert and Lapkiewicz, Radek and Zeilinger, Anton},
  journal={Physical review letters},
  volume={116},
  number={9},
  pages={090405},
  year={2016},
  publisher={APS},
  url={https://doi.org/10.1103/PhysRevLett.116.090405}
}

@article{karniadakis2021physics,
  title={Physics-informed machine learning},
  author={Karniadakis, George Em and Kevrekidis, Ioannis G and Lu, Lu and Perdikaris, Paris and Wang, Sifan and Yang, Liu},
  journal={Nature Reviews Physics},
  volume={3},
  number={6},
  pages={422--440},
  year={2021},
  publisher={Nature Publishing Group UK London},
  url={https://doi.org/10.1038/s42254-021-00314-5}
}

@article{nichols2019designing,
  author  = {Nichols, Rosanna and Mineh, Lana and Rubio, Jes{\'u}s and Matthews, Jonathan C. F. and Knott, Paul A.},
  title   = {Designing Quantum Experiments with a Genetic Algorithm},
  journal = {Quantum Science and Technology},
  volume  = {4},
  number  = {4},
  pages   = {045012},
  year    = {2019},
  doi     = {10.1088/2058-9565/ab4d89}
}

@article{knott2016search,
  title={A search algorithm for quantum state engineering and metrology},
  author={Knott, PA},
  journal={New Journal of Physics},
  volume={18},
  number={7},
  pages={073033},
  year={2016},
  publisher={IOP Publishing},
  url={https://doi.org/10.1088/1367-2630/18/7/073033}
}

@article{ruiz2026neural,
  title={Neural surrogates for designing gravitational wave detectors},
  author={Ruiz-Gonzalez, Carlos and Arlt, S{\"o}ren and Lehner, Sebastian and Berzins, Arturs and Drori, Yehonathan and Adhikari, Rana X and Brandstetter, Johannes and Krenn, Mario},
  journal={Machine Learning: Science and Technology},
  volume={7},
  number={5},
  pages={055010},
  year={2026},
  publisher={IOP Publishing},
  url={https://doi.org/10.1088/2632-2153/ae9de9}
}

@article{ma2021deep,
  title={Deep learning for the design of photonic structures},
  author={Ma, Wei and Liu, Zhaocheng and Kudyshev, Zhaxylyk A and Boltasseva, Alexandra and Cai, Wenshan and Liu, Yongmin},
  journal={Nature photonics},
  volume={15},
  number={2},
  pages={77--90},
  year={2021},
  publisher={Nature Publishing Group UK London},
  url={https://doi.org/10.1038/s41566-020-0685-y}
}

@article{peano2021rapid,
  title={Rapid exploration of topological band structures using deep learning},
  author={Peano, Vittorio and Sapper, Florian and Marquardt, Florian},
  journal={Physical Review X},
  volume={11},
  number={2},
  pages={021052},
  year={2021},
  publisher={APS},
  url={https://doi.org/10.1103/PhysRevX.11.021052}
}

@article{ruiz2023digital,
  author  = {Ruiz-Gonzalez, Carlos and Arlt, S{\"o}ren and Petermann, Jan and Sayyad, Sharareh and Jaouni, Tareq and Karimi, Ebrahim and Tischler, Nora and Gu, Xuemei and Krenn, Mario},
  title   = {Digital Discovery of 100 Diverse Quantum Experiments with {PyTheus}},
  journal = {Quantum},
  volume  = {7},
  pages   = {1204},
  year    = {2023},
  doi     = {10.22331/q-2023-12-12-1204}
}

@article{bao2023very,
  title={Very-large-scale integrated quantum graph photonics},
  author={Bao, Jueming and Fu, Zhaorong and Pramanik, Tanumoy and Mao, Jun and Chi, Yulin and Cao, Yingkang and Zhai, Chonghao and Mao, Yifei and Dai, Tianxiang and Chen, Xiaojiong and others},
  journal={Nature Photonics},
  volume={17},
  number={7},
  pages={573--581},
  year={2023},
  publisher={Nature Publishing Group UK London},
  url={https://doi.org/10.1038/s41566-023-01187-z}
}

@article{qian2023multiphoton,
  title={Multiphoton non-local quantum interference controlled by an undetected photon},
  author={Qian, Kaiyi and Wang, Kai and Chen, Leizhen and Hou, Zhaohua and Krenn, Mario and Zhu, Shining and Ma, Xiao-song},
  journal={Nature Communications},
  volume={14},
  number={1},
  pages={1480},
  year={2023},
  publisher={Nature Publishing Group UK London},
  url={https://doi.org/10.1038/s41467-023-37228-y}
}

@article{feng2023chip,
  title={On-chip quantum interference between the origins of a multi-photon state},
  author={Feng, Lan-Tian and Zhang, Ming and Liu, Di and Cheng, Yu-Jie and Guo, Guo-Ping and Dai, Dao-Xin and Guo, Guang-Can and Krenn, Mario and Ren, Xi-Feng},
  journal={Optica},
  volume={10},
  number={1},
  pages={105--109},
  year={2023},
  publisher={Optica Publishing Group}
}

@article{wang2024entangling,
  title={Entangling independent particles by path identity},
  author={Wang, Kai and Hou, Zhaohua and Qian, Kaiyi and Chen, Leizhen and Krenn, Mario and Zhu, Shining and Ma, Xiao-Song},
  journal={Physical Review Letters},
  volume={133},
  number={23},
  pages={233601},
  year={2024},
  publisher={APS},
  url={https://doi.org/10.1103/PhysRevLett.133.233601}
}

@article{krenn2017quantum,
  author  = {Krenn, Mario and Gu, Xuemei and Zeilinger, Anton},
  title   = {Quantum Experiments and Graphs: Multiparty States as Coherent Superpositions of Perfect Matchings},
  journal = {Physical Review Letters},
  volume  = {119},
  number  = {24},
  pages   = {240403},
  year    = {2017},
  doi     = {10.1103/PhysRevLett.119.240403}
}

@article{gu2019quantum,
  author  = {Gu, Xuemei and Erhard, Manuel and Zeilinger, Anton and Krenn, Mario},
  title   = {Quantum Experiments and Graphs {II}: Quantum Interference, Computation, and State Generation},
  journal = {Proceedings of the National Academy of Sciences},
  volume  = {116},
  number  = {10},
  pages   = {4147--4155},
  year    = {2019},
  doi     = {10.1073/pnas.1815884116}
}

@article{gu2019quantum3,
  title={Quantum experiments and graphs. III. High-dimensional and multiparticle entanglement},
  author={Gu, Xuemei and Chen, Lijun and Zeilinger, Anton and Krenn, Mario},
  journal={Physical Review A},
  volume={99},
  number={3},
  pages={032338},
  year={2019},
  publisher={APS},
  url={https://doi.org/10.1103/PhysRevA.99.032338}
}

@article{krenn2021conceptual,
  author  = {Krenn, Mario and Kottmann, Jakob S. and Tischler, Nora and Aspuru-Guzik, Al{\'a}n},
  title   = {Conceptual Understanding through Efficient Automated Design of Quantum Optical Experiments},
  journal = {Physical Review X},
  volume  = {11},
  number  = {3},
  pages   = {031044},
  year    = {2021},
  doi     = {10.1103/PhysRevX.11.031044}
}

@article{jaouni2024deep,
  author  = {Jaouni, Tareq and Arlt, S{\"o}ren and Ruiz-Gonzalez, Carlos and Karimi, Ebrahim and Gu, Xuemei and Krenn, Mario},
  title   = {Deep Quantum Graph Dreaming: Deciphering Neural Network Insights into Quantum Experiments},
  journal = {Machine Learning: Science and Technology},
  volume  = {5},
  number  = {1},
  pages   = {015029},
  year    = {2024},
  doi     = {10.1088/2632-2153/ad2628}
}

@inproceedings{kileel2019expressive,
  author    = {Kileel, Joe and Trager, Matthew and Bruna, Joan},
  title     = {On the Expressive Power of Deep Polynomial Neural Networks},
  booktitle = {Advances in Neural Information Processing Systems},
  volume    = {32},
  pages     = {10310--10319},
  year      = {2019},
  publisher = {Curran Associates, Inc.},
  url       = {https://proceedings.neurips.cc/paper/2019/hash/a0dc078ca0d99b5ebb465a9f1cad54ba-Abstract.html}
}

@article{kubjas2024geometry,
  title={Geometry of polynomial neural networks},
  author={Kubjas, Kaie and Li, Jiayi and Wiesmann, Maximilian},
  journal={Algebraic Statistics},
  volume={15},
  number={2},
  pages={295--328},
  year={2024},
  publisher={Mathematical Sciences Publishers},
  url={https://doi.org/10.2140/astat.2024.15.295}
}

@article{chrysos2021deep,
  author  = {Chrysos, Grigorios G. and Moschoglou, Stylianos and Bouritsas, Giorgos and Deng, Jiankang and Panagakis, Yannis and Zafeiriou, Stefanos},
  title   = {Deep Polynomial Neural Networks},
  journal = {IEEE Transactions on Pattern Analysis and Machine Intelligence},
  volume  = {44},
  number  = {8},
  pages   = {4021--4034},
  year    = {2022},
  doi     = {10.1109/TPAMI.2021.3058891}
}

@article{pizarroso2022neuralsens,
  author  = {Pizarroso, Jaime and Portela, Jos{\'e} and Mu{\~n}oz, Antonio},
  title   = {{NeuralSens}: Sensitivity Analysis of Neural Networks},
  journal = {Journal of Statistical Software},
  volume  = {102},
  number  = {7},
  pages   = {1--36},
  year    = {2022},
  doi     = {10.18637/jss.v102.i07}
}

@article{erhard2020advances,
  author  = {Erhard, Manuel and Krenn, Mario and Zeilinger, Anton},
  title   = {Advances in High-Dimensional Quantum Entanglement},
  journal = {Nature Reviews Physics},
  volume  = {2},
  number  = {7},
  pages   = {365--381},
  year    = {2020},
  doi     = {10.1038/s42254-020-0193-5}
}

@misc{hendrycks2016gaussian,
  author        = {Hendrycks, Dan and Gimpel, Kevin},
  title         = {Gaussian Error Linear Units ({GELUs})},
  year          = {2016},
  eprint        = {1606.08415},
  archivePrefix = {arXiv},
  primaryClass  = {cs.LG}
}

@article{olearyroseberry2024derivative,
  author  = {O'Leary-Roseberry, Thomas and Chen, Peng and Villa, Umberto and Ghattas, Omar},
  title   = {Derivative-Informed Neural Operator: An Efficient Framework for High-Dimensional Parametric Derivative Learning},
  journal = {Journal of Computational Physics},
  volume  = {496},
  pages   = {112555},
  year    = {2024},
  doi     = {10.1016/j.jcp.2023.112555}
}

@article{luo2025efficient,
  author  = {Luo, Dingcheng and O'Leary-Roseberry, Thomas and Chen, Peng and Ghattas, Omar},
  title   = {Efficient {PDE}-Constrained Optimization under High-Dimensional Uncertainty Using Derivative-Informed Neural Operators},
  journal = {SIAM Journal on Scientific Computing},
  volume  = {47},
  number  = {4},
  pages   = {C899--C931},
  year    = {2025},
  doi     = {10.1137/23M157956X}
}

@inproceedings{greydanus2019hamiltonian,
  author    = {Greydanus, Samuel and Dzamba, Misko and Yosinski, Jason},
  title     = {Hamiltonian Neural Networks},
  booktitle = {Advances in Neural Information Processing Systems},
  volume    = {32},
  year      = {2019},
  publisher = {Curran Associates, Inc.},
  url       = {https://proceedings.neurips.cc/paper/2019/hash/26cd8ecadce0d4efd6cc8a8725cbd1f8-Abstract.html}
}

@article{lee2022complex,
  author  = {Lee, Chi Yan and Hasegawa, Hideyuki and Gao, Shangce},
  title   = {Complex-Valued Neural Networks: A Comprehensive Survey},
  journal = {IEEE/CAA Journal of Automatica Sinica},
  volume  = {9},
  number  = {8},
  pages   = {1406--1426},
  year    = {2022},
  doi     = {10.1109/JAS.2022.105743}
}

@article{meng2025physics,
  title={When physics meets machine learning: A survey of physics-informed machine learning},
  author={Meng, Chuizheng and Griesemer, Sam and Cao, Defu and Seo, Sungyong and Liu, Yan},
  journal={Machine Learning for Computational Science and Engineering},
  volume={1},
  number={1},
  pages={20},
  year={2025},
  publisher={Springer},
  url={https://doi.org/10.1007/s44379-025-00016-0}
}

@article{peurifoy2018nanophotonic,
  title={Nanophotonic particle simulation and inverse design using artificial neural networks},
  author={Peurifoy, John and Shen, Yichen and Jing, Li and Yang, Yi and Cano-Renteria, Fidel and DeLacy, Brendan G and Joannopoulos, John D and Tegmark, Max and Solja{\v{c}}i{\'c}, Marin},
  journal={Science advances},
  volume={4},
  number={6},
  pages={eaar4206},
  year={2018},
  publisher={American Association for the Advancement of Science},
  url={https://doi.org/10.1126/sciadv.aar4206}
}

@inproceedings{cranmer2020lagrangian,
  title     = {{Lagrangian Neural Networks}},
  author    = {Cranmer, Miles and Greydanus, Sam and Hoyer, Stephan and Battaglia, Peter and Spergel, David and Ho, Shirley},
  booktitle = {ICLR 2020 Workshops: DeepDiffEq},
  year      = {2020},
  url       = {https://mlanthology.org/iclrw/2020/cranmer2020iclrw-lagrangian/}
}

@inproceedings{lee2019set,
  title={Set transformer: A framework for attention-based permutation-invariant neural networks},
  author={Lee, Juho and Lee, Yoonho and Kim, Jungtaek and Kosiorek, Adam and Choi, Seungjin and Teh, Yee Whye},
  booktitle={International conference on machine learning},
  pages={3744--3753},
  year={2019},
  organization={PMLR},
  url={https://proceedings.mlr.press/v97/lee19d.html}
}

@article{vaswani2017attention,
  title={Attention is all you need},
  author={Vaswani, Ashish and Shazeer, Noam and Parmar, Niki and Uszkoreit, Jakob and Jones, Llion and Gomez, Aidan N and Kaiser, {\L}ukasz and Polosukhin, Illia},
  journal={Advances in neural information processing systems},
  volume={30},
  year={2017},
   url = {https://proceedings.neurips.cc/paper_files/paper/2017/file/3f5ee243547dee91fbd053c1c4a845aa-Paper.pdf}
}


\clearpage
\onecolumngrid
\appendix
\section{Additional Jacobian-Based Sensitivity Comparisons}
\label{app:perfect-matching-derivation}


\begin{figure}[!ht]
    \centering
    \includegraphics[width=1\textwidth]{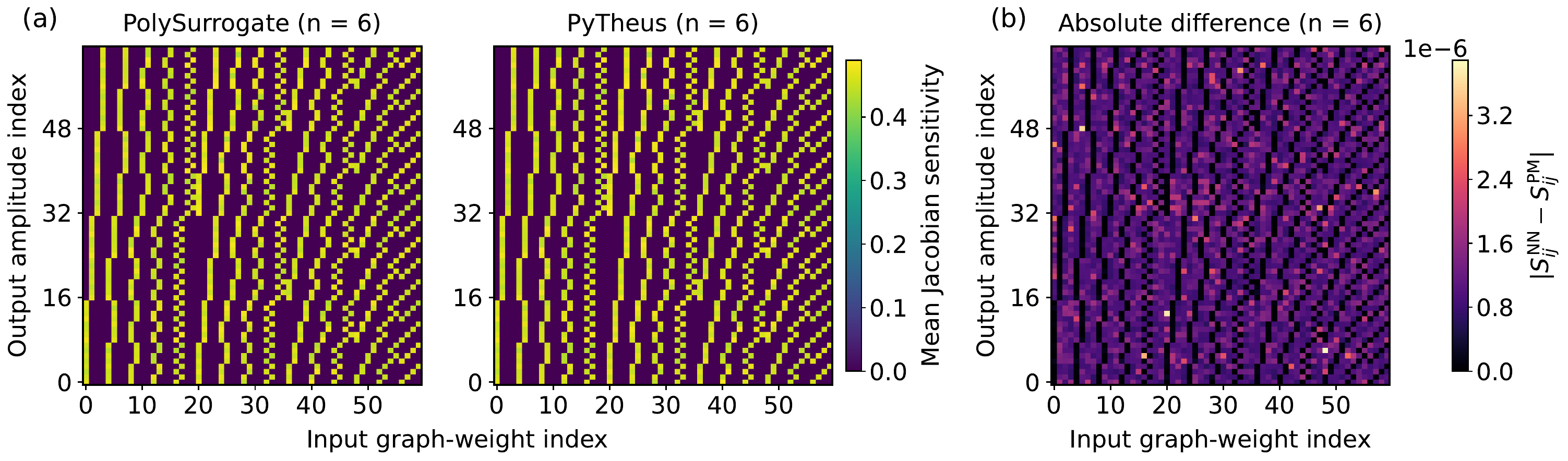}
     \caption{\textbf{Jacobian-based sensitivity comparison for the \(n=6\) graph system.}
\textbf{(a)} Mean absolute sensitivities from the PNN (left) and the analytical perfect-matching calculation (right), evaluated on the same \(1000\) sampled graphs. Rows represent output-state amplitudes and columns represent graph weights.
\textbf{(b)} Absolute element-wise difference between the two sensitivity matrices. Their matching patterns and small differences show that the PNN reproduces the mean absolute sensitivities of the output amplitudes to the graph weights.
}

    \label{fig:jacobian-sensitivity-6}
\end{figure}

\begin{figure}[!ht]
    \centering
    \includegraphics[width=1\textwidth]{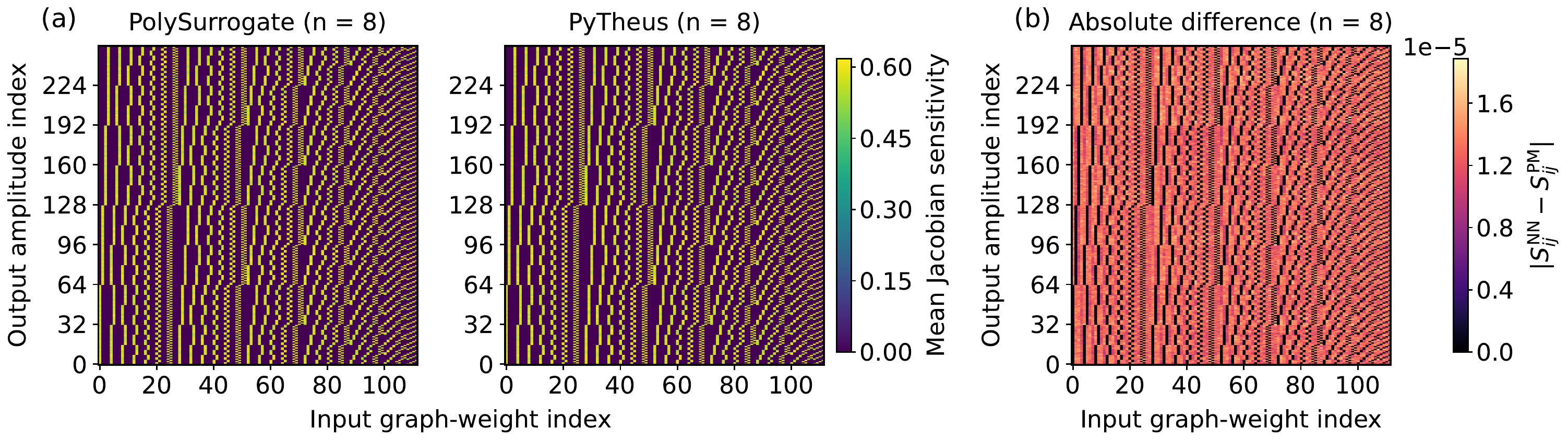}
     \caption{\textbf{Jacobian-based sensitivity comparison for the \(n=8\) graph system.}
\textbf{(a)} Mean absolute sensitivities from the PNN (left) and the analytical perfect-matching calculation (right), evaluated on the same \(1000\) sampled graphs. Rows represent output-state amplitudes and columns represent graph weights.
\textbf{(b)} Absolute element-wise difference between the two sensitivity matrices. Their matching patterns and small differences show that the PNN reproduces the mean absolute sensitivities of the output amplitudes to the graph weights.
}

    \label{fig:jacobian-sensitivity-8}
\end{figure}


\end{document}